\documentstyle{article}%
\begin{document}

\title{Protein secondary structure prediction based on quintuplets}
\author{Wei-Mou Zheng\\
{\it Institute of Theoretical Physics, Academia Sinica, Beijing
100080, China, }\\ and\\ {\it Bejing Genomics Institute, Academia
Sinica, Beijing 101300, China } } \date{}\maketitle
\begin{abstract}
Simple hidden Markov models are proposed for predicting secondary
structure of a protein from its amino acid sequence. Since the
length of protein conformation segments varies in a narrow range,
we ignore the duration effect of length distribution, and focus on
inclusion of short range correlations of residues and of
conformation states in the models. Conformation-independent and
-dependent amino acid coarse-graining schemes are designed for the
models by means of proper mutual information. We compare models of
different level of complexity, and establish a practical model
with a high prediction accuracy.
\end{abstract}

\leftline{PACS number(s): 87.10.+e,02.50.-r}%
\bigskip

\section{Introduction}

Methods for predicting the secondary structure of a protein from
its amino acid sequence have been developed for 3 decades. Besides
neural network models and nearest-neighbor methods, the
Chou-Fasman/GOR statistical method is well-established and
commonly used. In 1974, assuming an oversimplified independency to
cope with the large size 20 of the amino acid alphabets at a small
size of database, Chou and Fasman (1974) derived a table of
propensity for a particular residue to be in a given secondary
structure state. By combining with a set of rules, the protein
secondary structure was predicted using this propensity. Later, in
the first version of the GOR program (Garnier, Osguthorpe, and
Robson, 1978), the state of a single residue $a_i$ was predicted
according to a window from $i-8$ to $i+8$ surrounding the residue.
Unlike Chou-Fasman which assumes that each amino acid individually
influences its own secondary structure state, GOR takes into
account the influence of the amino acids flanking the central residue
on the central residue state by deriving an information
score from the weight matrix describing 17 individual amino acid
frequencies at sites $i+k$ with $-8\leq k\leq +8$. By using a
single weight matrix, the correlation among amino acids within the
window was still ignored. In the later version GOR III (Gibrat,
Garnier, and Robson, 1987), instead of single weight matrix for
every structure state, 20 
weight matrices, each of which corresponds to a specific type of
the central residue, were used. These conditional weight matrices
take the pair correlation between the central residue and a
flanking one into account. In the most recent version of GOR (GOR
IV, Garnier, Gibrat, and Robson, 1996), all pairwise combinations
of amino acids in the flanking region were included.

Hidden Markov models (HMMs) (Rabiner, 1989) have been applied
to molecular biology, in particular in gene-finding. There is a
constant tendency in developing HMMs for protein structure
prediction (Asai et al.,1993;
Stultz et al., 1993; Karplus et al., 1997; Camproux et al. 1999).
A probabilistic approach similar to the gene
finder Genscan has been developped for protein secondary structure
prediction without using sliding windows (Schmidler, Liu and Brutlag,
2000). In terms of Bayesian segmentation,
the mothod integrated explicit models for secondary structure
classes helices, sheets and loops with other observed structure
aspects such as segment capping signals and length distributions,
and reached an accuracy comparable to GOR.

Compared with DNA sequences, protein sequences are generally short,
and their amino acid alphabet is of a large size 20. The range of
lengths of secondary structure segments is rather small. The effect
of duration might play a less important role. Here we develop a
simple hidden Markov model with higher order correlations included
for the secondary structure prediction. We propose several schemes
for the amino acid alphabet reduction in order to incorporate residue
correlation in the model. While the model is much simpler than the
Bayesian segmentation model, its performance is still competitive.

\section{Methods}

A simplified version of the model can be constructed in the frame
of the Chou-Fasman propensity scheme. We shall start with this model
to explain several key points, and then discuss more realistic models.

In the Chou-Fasman approach, discriminant thresholds and
post-prediction filtering rules are required. They can be avoided in
a full probabilistic model.

As in the most methods, we consider 3 states $\{h, e, c\}$ generated
from the 8 states of Kabsch and Sander (1983) by the coarse-graining
$H,G,I\to h$, $E\to e$ and $X,T,S,B\to c$. Let $R\equiv R_{1:n}
=R_1R_2\ldots R_n$ be
a sequence of $n$ amino acid residues, and its corresponding secondary
structure sequence be $S =S_1S_2\ldots S_n$.
The structure prediction
is the mapping from $R$ to $S$. The main restriction to a
structure sequence is that the shortest length of the consecutive
state $h$ must be 3, and that of $e$ be 2. To cope with this
restriction, we use triplet states instead of the 3 single states $c$,
$e$ and $h$. In the total 27 triplets, only 19 are legitimate. The
forbidden 8 are of the type $\bar e e\bar e$ or $\bar h h\bar h$, where
$\bar e$ indicates a non-$e$, i.e. either $c$ or $h$, and the meaning
of $\bar h$ is analogous. The first order Markov model for the triplets
is the third order Markov model for the original mono-states. Any of
the 19 triplets can only transit either to 3 or to 1 state. That is, the
transition matrix is rather sparse. The 19 triplets and their transited
states are listed in Table~1.

Denoting by $\sigma_i$ the triplet states, we may translate $S_{1:n}$ to
$\Sigma_{2:n-1}= \sigma_2\sigma_3\ldots \sigma_{n-1}$ of length $n-2$.
(Note that the subscripts $i$ for $\sigma_i$ are from 2 to $n-1$ for
convenience.) The Markov process for $\sigma$ is characterized by the
set of probabilities for initial states
\begin{equation}
\pi (l)={\rm Prob}(\sigma_2 =l),
\label{pi}\end{equation}
and the transition rates
\begin{equation}
T_{kl} ={\rm Prob}( \sigma_i=l|\sigma_{i-1}=k)=
T(\sigma_{i-1},\sigma_i). \label{tr}\end{equation} Sequence $R$ is
then related to $\Sigma$ or $S$ by the emission probabilities
\begin{equation}
P(x|\delta )= {\rm Prob}( R_i=x |\sigma_i=\delta ),
\label{em}\end{equation}
which generate $S_{2:n-1}$. Extra probabilities
\begin{equation}
Q(x|\delta )= {\rm Prob}( R_j=x |\sigma_{j\pm 1} =\delta ), j\in \{1,n\}
\label{em}\end{equation}
are needed for generating $R_1$ and $R_n$.

In this model the probability for the state sequence $S$ or $\Sigma$ is
\begin{equation}
P(\Sigma ) =\pi (\sigma_2) \prod_{i=2}^{n-2} T(\sigma_i, \sigma_{i+1}),
\end{equation}
and the likelihood for $R$ to be at $S$ is
\begin{equation}
P(R|\Sigma ) =Q(S_1|\sigma_2)Q(S_n|\sigma_{n-1})\prod_{i=2}^{n-1} P(S_i|\sigma_i ).
\end{equation}
The joint probability is then
\begin{equation}
P(R,\Sigma )= P(R,S) = P(R|\Sigma )P(\Sigma ).
\end{equation}
The predicted structure is infered as
\begin{equation}
\Sigma^* = {\rm arg\,max}_{\Sigma} P(\Sigma |R) \propto
{\rm arg\,max}_{\Sigma} P(R,\Sigma ).
\end{equation}
By means of the recursion relation
\begin{eqnarray}
\Gamma_2(\sigma )&=& Q(S_1|\sigma )\pi (\sigma )P(R_2|\sigma ), \\
\Gamma_i(\sigma )&=& {\rm max}_{\delta} \Gamma_{i-1}(\delta )
T(\delta ,\sigma )P(R_i|\sigma ),\qquad 2<i<n, \\
\Gamma &\equiv &\Gamma_n ={\rm max}_{\delta}
\Gamma_{n-1}(\delta )Q(S_n|\delta ),
\end{eqnarray}
and by recording the pre-state leading to the maximal $\Gamma_i(\sigma )$,
the `best' path $\Sigma^*$ can be traced back from the last
$\sigma^*_{n-1}$. This is the so-called Viterbi algorithm for dynamic
programming.

 According to the standard forward-backward algorithm for HMMs, the
forward and backward variables $A_i$ and $B_i$ may be defined
as follows.
\begin{eqnarray}
A_2(\sigma )&=& Q(S_1|\sigma )\pi (\sigma )P(R_2|\sigma ), \\
A_i(\sigma )&=& \sum_{\delta} A_{i-1}(\delta ) T(\delta ,\sigma
)P(R_i|\sigma ),\qquad 2<i<n.
\end{eqnarray}
Similarly,
\begin{eqnarray}
B_{n-1}(\sigma )&=& Q(S_n|\sigma ), \\
B_i(\sigma )&=& \sum_{\delta} B_{i+1}(\delta )
T(\sigma ,\delta )P(R_{i+1}|\delta ),\qquad 2\leq i<n-1. \\
\end{eqnarray}
It can be seen that for non-ending $i$, $A_i(\sigma )= {\rm Prob} (R_{1:i},
\sigma_i=\sigma )$, $B_i(\sigma )= {\rm Prob}(R_{i+1:n}|\sigma_i=
\sigma )$, and the partition function
\begin{eqnarray}
Z &\equiv& \sum_\Sigma P(R,\Sigma ) =\sum_\sigma A_i(\sigma )B_i(\sigma )\\
&=& \sum_\sigma A_i(\sigma )B_i(\sigma ),\qquad 2\leq i\leq n-1.
\end{eqnarray}
Denoting $\sigma^{(0)}_i$ the center $S_i$ of the triplet $\sigma_i=S_{i-1}S_i
S_{i+1}$, and introducing the characteristic function for $z\in \{c, e, h\}$
\begin{equation}
\delta (\sigma_i ,z) =\left\{
\begin{array}{ll}
1, &{\rm if } \sigma^{(0)}_i=z,\\
0,& {\rm otherwise,}
\end{array} \right.
\end{equation}
we may infer single residue state from the marginal posterior
\begin{equation}
{\rm Prob} (S_i=z|R ) \propto \sum_\sigma A_i(\sigma )B_i(\sigma )
\delta (\sigma ,z).
\end{equation}
This is the Baum-Welch algorithm for single residue states.

So far, only the correlation of conformation states has been
considered. Residue triple will involve $20^3 =8000$ parameters
for each state $\sigma$. To avoid large training sets and model
overfitting, a reduced amino acid alphabet is desired. For
example, the reduction of 20 amino acids to 3 classes leads to
only 27 combinations. However, there are as many as
$$ \frac 1{3!}\sum_{j=0}^3 C_3^j (-1)^j (3-j)^{20} \approx 5.8\times 10^8$$
ways of clustering 20 amino acids into 3 classes (Duran and Odell,
1974). For a given clustering $\{a_i\}_{i=0}^{19}\to \{b_j\},
b_j\in \{0, 1, 2\}$, denoting by $\rho_i=r_{i-1}r_ir_{i+1}$ the
reduced residue triple corresponding to state $\sigma_i$, we may
calculate the mutual information between the reduced residue
triple $\rho$ and the triple state $\sigma$:
\begin{equation}
I(\rho ,\sigma )=H[\rho ] +H[\sigma ]-H[\rho ,\sigma ],
\label{i1}\end{equation}
where $H[x]$ is the Shannon entropy of $x$. (If the clustering
independent $H[\sigma ]$ is ignored, $I$ would become the conditional
entropy $H[\sigma |\rho ]$.) The best clustering may be determined by
maximizing the objective function $I(\rho ,\sigma )$. The replacement of
the above $P(R_i|\sigma_i)$ with $P(\rho_i|\sigma_i)$ leads to a version
which takes some residue correlation into account.

A more realistic model uses quintuplets. Among the total $3^5=243$
quintuplets of the conformation states, only 75 are legitimate.
Exclusion of 7 rare ones ($eceeh$, $hceeh$, $heece$, $heech$,
$heeeh$, $heehh$, and $hheeh$) further reduces the total number of
states into 68, which are listed in Table~2. We shall still use
the same notation $\sigma_i$ for these 68 conformation states. To
take residue quintuplet correlation into account, we substitute
the central residue score $P(R_i|\sigma_i)$ with $P(R_i|\sigma_i,
r_{i-2}r_{i-1}r_{i+1}r_{i+2})$, where $r_i$ stands for reduced
residue classes. More words need to be said about the amino acid
clustering. We have observed the fact that counts of $ccccc$,
$eeeee$ and $hhhhh$ in databases are dominant over those of
remaining 65. The various propensities of amino acid residues to
different conformations imply that amino acid clustering should
depend on conformations. We want to cluster amino acids separately
for each conformation. For this purpose, for example, to find the
best clustering at conformation $c$, we collect a subset of
residue quintuplets whose conformation is $ccccc$. Denote by $R_0$
the central residue of a residue quintuplet, and by
$r_{-2}r_{-1}r_1r_2$ the reduced classes of the other 4 residues.
Taking the mutual information $I(R_0, r_{-2}r_{-1}r_1r_2)$ as the
objective function, we determine the best clustering at $c$. To
find at which position the residue depends most strongly on
others, we calculate mutual information for nonreduced residue
placed at different positions of a residue quintuplet. While the
largest $I$ is found when the nonreduced residue is at the center
for conformation $e$, the position of the nonreduced residue which
gives the largest $I$ is the second position for conformation $c$,
and is the fifth for $h$. However, for either $c$ or $h$, the
largest $I$ is still very close to $I(R_0, r_{-2}r_{-1}r_1r_2)$.
The mutual information excesses with respect to $I(R_0,
r_{-2}r_{-1}r_1r_2)$ for different positions at conformations $c$,
$e$ and $h$ are listed in Table~3. Thus, for simplicity, we always
place the nonreduced residue at the center for all conformations.
We then calculate refined residue scores from the
conformation-dependent clustering. For example,
\begin{equation}
P(R_i|\sigma_i=ccchh, r_{i-2}r_{i-1}r_{i+1}r_{i+2}) \to
P(R_i|\sigma_i, r_{i-2}^cr_{i-1}^cr_{i+1}^hr_{i+2}^h),
\end{equation}
where the superscript of $r_i$ indicates its conformation. The
whole procedure for the dynamic programming remains almost the
same, except for some care when dealing with two more end sites.

\section{Result}

We create a nonredundant set of 1612 non-membrane proteins for
training parameters from PDB\_SELECT (Hobohm and Sander, 1994)
with amino acid identity less than 25\% issued on 25 September of
2001. The secondary structure for these sequences are taken from
DSSP database (Kabsch and Sander, 1983). As mentioned above, the
eight states of DSSP are coarse-grained into 3 states: $h$, $e$
and $c$. This learning set contains 268031 residues with known
conformations, among which 94415 are $h$, 56510 are $e$, and
117106 are $c$. The size of the learning set is reasonable for
training our parameters. There are 296 unknown residues. We add an
extra `unknown' amino acid category called $X$ to the 20 known
ones.

In order to assess the accuracy of our approach, we use the following 2
test sets: Sets 1 and 2. A set of 124 nonhomologous proteins is
created from the representative database of Rost and Sander (1993)
by removing subunits A and B of hemagglutinin 3hmg, which are
designated as membrane protein by SCOP (Murzin et al, 1995). The
124 sequences and the learning set are not independent of each
other according to HSSP database (Dodge, Schneider and Sander,
1998). That is, some proteins of the 124 sequences are homologous
with certain proteins in the learning set. Removing the homologous
proteins from the 124 sequences and 5 seuqences with unknown amino
acid segments longer than 6, we construct Set 1 of 76
proteins. Nonredundant 34 proteins
with known structures of the CASP4 database issued in December of
2000 are taken as Set 2 (CASP4, 2000).

\subsection{Amino acid clustering}
The first method for clustering uses the mutual information $I(\rho ,
\sigma )$ between the reduced oligo-peptide $\rho$ and its corresponding
conformation $\sigma$. Setting the number of reduced classes at 3, 4 and
5, and fixing $\sigma$ to be triplets, we find the
conformation-independent clustering of amino acids as shown in Table~4.
Roughly speaking, class 0 is the hydrophobic, and is the same for the
all 3 clusterings (except for the `unknown' $X$ in the clustering into 5).
Increasing the number of classes from 3 to 4, 5 results in new classes
forming by single special amino acids P(Pro) and G(Gly). The similar
clustering may be conducted in terms of quintuplets, but the results
do not coincide with those from triplets.

The second method for clustering is conformation-dependent. We
collect residue quintuplets of conformations $ccccc$, $eeeee$ and
$hhhhh$ separately. The objective function for clustering now is
the mutual information $I(R_0, r_{-2}r_{-1}r_1r_2|\sigma )$ with
conformation $\sigma$ fixed. The results of clustering into 3 and
4 classes are listed in Table~5. Indeed, the cluster patterns for
different conformations are quite dissimilar.

\subsection{Secondary structure prediction}
We shall index different models by $N_s$-$N_a$, where $N_s$ and
$N_a$ are the numbers of conformation states and residue
combinations, respectively. For example, the simplest model is
model 19-21, which uses $P(R_i|S_{i-1}S_iS_{i+1})$ to score
residues. We may also use the propensity scores
$P(R_i|S_{i-1}S_iS_{i+1})/P(R_i)$, which result in an extra factor
$\prod_{i=0}^n P(R_i)$. Since the factor is independent of
conformation sequences, it brings no new effect. When the scores
are replaced by $P(r_{i-1}r_ir_{i+1}|S_{i-1}S_iS_{i+1})$ with
$r_i$ being the reduced 3-class residues, we have model 19-$3
\times 3\times 3$. The first and last sites of the predicted
conformation of any sequences are always set at conformation $c$.
For model 19-21 we determine the best conformation sequence as
whole by the Viterbi algorithm and single residue conformation
states by the Baum-Welch algorithm. To assess prediction methods,
we calculate for each conformation the sensitivity $s_n$ and
specificity $s_p$
\begin{equation}
s_n= \frac{TP}{TP+FN},\quad s_p= \frac{TP}{TP+FP},
\end{equation}
where $TP$, $FP$ and $FN$ are site counts of the `true positive',
`false positive' and `true negative' with respect to the observed
real conformation. The results of model 19-21 are listed in
Table~6, where the total sensitivity $Q_3$ for all conformations
is also given. It is clearly seen that the inference from the
Baum-Welch marginal posterior is significantly superior to that
from the Viterbi algorithm in $Q_3$ value.

The next examined model uses reduced residue triplets. Reducing amino
acids into 3 and 4 classes, we have models 19-27 and 19-64, respectively.
The prediction accuracies of these two models are also listed in Tabel~6.
We see that the inclusion of residue correlation dramatically improves
the prediction accuracy.

The remaining part of Table~6 shows the prediction accuracies of
quintuplet models. We examine the models on both test Sets 1 and
2. For all the models we take
$P(R_i|r_{i-2}r_{i-1}r_{i+1}r_{i+2},\sigma_i )$ as the residue
scores. We first compare conformation-independent with
conformation-dependent clustering of amino acids. We find that the
conformation-dependent clustering gains about 1 percent in the
prediction accuracy for model 68-81$\times$21. The
conformation-independent clustering is then not considered later
on. Model 68-256$\times$21 contains more information about
correlated residues, and has a better performance. The accuracies
obtained on Set 1 are generally higher than those on Set 2.
Besides the popular predictor GOR IV, there is another secondary
structure predictor SSP (Solovyev and Salamov, 1991, 1994) based
on discriminant analysis using single sequence. To compare with
them, their accuracies on the same test sets are also listed in
the table.

\section{Discussions}

We have presented simple hidden Markov models to predict secondary
structure using single protein sequence. The hidden sequence is
generated by a Markov process of multi-site conformation states.
Considering that structure segments of proteins are generally
short, we have ignored the duration effect, and focused on short
range correlations. We proposed several schemes for
coarse-graining the amino acid alphabet in order to include
multi-residue correlation. Such reduction has been used in the
Bayesian segmentation of protein secondary structure (Schmidler,
et. al., 2000). However, here we derived the coarse-graining
schemes specially for scoring residues to fit conformations. We
have discussed only the principle of taking proper mutual
information as an objective function for clustering, but did not
exhaust all possibilities for clustering. For example, to diminish
parameters, one may consider residue triplets at conformation
quintuplet states. Another possibility is to use quartuples for
both residues and conformations. One can also cluster quintuplet
conformation states to less states.

There are rooms for further improvement of our approach. Simple
weights (3 values) may be introduced to adjust residue scores
according to its single-site conformation. Moreover, we may divide
a training set into several, say 2, subsets according to residue
statistics. Again, for this purpose the coarse-graining schemes
help. The two subsets are then used separately for training to get
refined models. We first classify a query sequence into one of the
two categories, and then apply to it the corresponding refined
model. We have tested this on the simple triplet models. The
primitive result of up to 2\% in accuracy improvement is
encouraging. This, and the protein family recognition using
multiple reduced amino acid residues, are under study.

\begin{quotation}
{This work was supported in part by the Special Funds for Major
National Basic Research Projects and the National Natural Science
Foundation of China.}
\end{quotation}

\newpage
Table 1. Triplet states and their transited states.\\

\begin{tabular}{|r|c|c||r|c|c|}
\hline
\multicolumn{2}{|c|}{State}&Transited &\multicolumn{2}{|c|}{State}&Transited \\
\hline
 0& CCC&  0, 1, 2& 10& EEH& 11\\
 1& CCE&  3& 11& EHH& 18\\
 2& CCH&  4& 12& HCC&  0, 1, 2\\
 3& CEE&  8, 9,10& 13& HCE&  3\\
 4& CHH& 18& 14& HCH&  4\\
 5& ECC&  0, 1, 2& 15& HEE&  8, 9,10\\
 6& ECE&  3& 16& HHC& 12,13,14\\
 7& ECH&  4& 17& HHE& 15\\
 8& EEC&  5, 6, 7& 18& HHH& 16,17,18\\
 9& EEE&  8, 9,10& &&\\ \hline
\end{tabular}\\

\bigskip
Table 2. 68 quintuplet states.\\

\begin{tabular}{|rc|rc|rc|rc|}
\hline
 0& CCCCC& 17& CHHHE& 34& EEECH& 51& HEECC\\
 1& CCCCE& 18& CHHHH& 35& EEEEC& 52& HEEEC\\
 2& CCCCH& 19& ECCCC& 36& EEEEE& 53& HEEEE\\
 3& CCCEE& 20& ECCCE& 37& EEEEH& 54& HHCCC\\
 4& CCCHH& 21& ECCCH& 38& EEEHH& 55& HHCCE\\
 5& CCEEC& 22& ECCEE& 39& EEHHH& 56& HHCCH\\
 6& CCEEE& 23& ECCHH& 40& EHHHC& 57& HHCEE\\
 7& CCEEH& 24& ECEEC& 41& EHHHE& 58& HHCHH\\
 8& CCHHH& 25& ECEEE& 42& EHHHH& 59& HHEEC\\
 9& CEECC& 26& ECHHH& 43& HCCCC& 60& HHEEE\\
10& CEECE& 27& EECCC& 44& HCCCE& 61& HHHCC\\
11& CEECH& 28& EECCE& 45& HCCCH& 62& HHHCE\\
12& CEEEC& 29& EECCH& 46& HCCEE& 63& HHHCH\\
13& CEEEE& 30& EECEE& 47& HCCHH& 64& HHHEE\\
14& CEEEH& 31& EECHH& 48& HCEEC& 65& HHHHC\\
15& CEEHH& 32& EEECC& 49& HCEEE& 66& HHHHE\\
16& CHHHC& 33& EEECE& 50& HCHHH& 67& HHHHH\\
\hline\end{tabular}\\

\bigskip
Table 3. Mutual information excesses ($\times 10^{-3}$) for
different positions of the nonreduced residue with respect to the
mutual information for the nonreduced residue at the quintuplet
center. The center is referred to
as position 3.\\

\begin{tabular}{|c|rrrr|}\hline
& \multicolumn{4}{|c|}{Position of the nonreduced}\\
Conformation& \multicolumn{1}{c}1& \multicolumn{1}{c}2
& \multicolumn{1}{c}4& \multicolumn{1}{c|}5\\
\hline
c& -19.5&   2.7&  -3.0&  -7.4\\
e& -46.1& -10.2& -15.5& -82.3\\
h& -49.6& -33.8&  -0.5&   1.8\\
\hline\end{tabular}\\
\newpage
Table 4. Conformation-independent clustering of amino acids into
3, 4 and 5 classes. \\

\begin{tabular}{|c|ccccccccccccccccccccc|}\hline
Amino Acid& A& V& C& D& E& F& G& H& I& W& K& L& M& N& Y& P& Q& R& S& T& X\\
\hline
3-class& 1& 0& 0& 2& 1& 0& 2& 1& 0& 0& 1& 0& 0& 2& 0& 2& 1& 1& 1& 1& 1\\
4-class& 1& 0& 0& 2& 1& 0& 2& 2& 0& 0& 1& 0& 0& 2& 0& 3& 1& 1& 2& 2& 1\\
5-class& 1& 0& 0& 2& 1& 0& 4& 2& 0& 0& 1& 0& 0& 2& 0& 3& 1& 1& 2& 2& 0\\
\hline\end{tabular}\\

\bigskip
Table 5. Conformation-dependent clustering of amino acids into
3 and 4 classes. \\

\begin{tabular}{|c|ccccccccccccccccccccc|}\hline
Amino Acid& A& V& C& D& E& F& G& H& I& W& K& L& M& N& Y& P& Q& R& S& T& X\\
\hline
c, 3-class& 1& 2& 0& 0& 1& 2& 0& 0& 2& 2& 1& 2& 2& 0& 2& 0& 1& 1& 1& 1& 2\\
e, 3-class& 0& 2& 0& 1& 1& 2& 0& 1& 2& 2& 1& 0& 0& 1& 2& 0& 1& 0& 1& 1& 1\\
h, 3-class& 0& 2& 0& 1& 1& 2& 1& 1& 2& 2& 1& 2& 2& 1& 2& 1& 1& 1& 1& 1& 0\\
\hline
c, 4-class& 1& 2& 2& 0& 1& 2& 3& 1& 2& 2& 1& 2& 2& 0& 2& 0& 1& 1& 1& 1& 2\\
e, 4-class& 0& 1& 2& 3& 3& 1& 2& 0& 1& 0& 3& 2& 2& 3& 1& 2& 3& 0& 0& 3& 2\\
h, 4-class& 1& 2& 1& 3& 0& 2& 3& 3& 2& 2& 0& 2& 2& 3& 1& 3& 3& 0& 3& 0& 1\\
\hline\end{tabular}\\

\bigskip Table 6. Accuracy of secondary structure predictions for
different models. Here, VI and BW stand for `Viterbi' and
`Baum-Welch' algorithms, respectively. For all models 68-k,
conformation-dependent reductions are used except for model
68-81$\times$21* where the single
conformation-independent reduction is used.\\

\begin{tabular}{|c|c|rrrrrrr|}\hline
Model&  Test set& $S_n^c$& $S_p^c$& $S_n^e$& $S_p^e$& $S_n^h$& $S_p^h$& $Q_3$\\
\hline
19-21, VI& 2& 48.69& 62.92& 7.94& 64.09& 84.17& 48.07& 53.24\\
19-21, BW& 2& 70.67& 62.27& 30.74& 54.55& 66.12& 60.11& 60.45\\
\hline
19-27, VI& 2& 62.50& 63.87& 56.02& 46.58& 55.82& 61.48& 58.63\\
19-64, VI& 2& 64.10& 65.31& 50.11& 47.59& 62.45& 63.06& 60.50\\
\hline
68-81$\times$21*, VI& 2& 60.12& 69.16& 44.54& 54.89& 75.48& 60.26& 62.53\\
\hline
68-81$\times$21, VI& 2&  62.01& 68.21& 44.26& 57.79& 76.51& 61.98& 63.64\\
68-81$\times$21, BW& 2&  74.51& 61.71& 47.18& 59.99& 61.52& 68.93& 63.83\\
\hline
68-81$\times$21, VI& 1&  70.92& 66.51& 54.27& 53.45& 62.00& 69.07& 64.46\\
68-81$\times$21, BW& 1&  71.01& 68.98& 52.97& 57.58& 67.37& 66.69& 66.00\\
\hline
68-256$\times$21, VI& 2& 65.35& 68.41& 54.73& 58.41& 70.11& 64.58& 64.86\\
68-256$\times$21, BW& 2& 70.90& 68.15& 57.60& 60.86& 68.85& 69.84& 67.30\\
\hline
68-256$\times$21, VI& 1& 68.34& 69.97& 55.34& 57.68& 70.12& 66.23& 66.18\\
68-256$\times$21, BW& 1& 73.22& 69.77& 57.40& 59.61& 67.20& 70.38& 67.90\\
\hline
GOR4&1& 79.3& 66.1& 54.7& 55.3& 63.3& 68.5& 66.2\\
SSP& 1& 59.2& 52.8& 69.0& 55.3& 67.0& 68.1& 60.0\\
GOR4& 2& 81.9& 62.0& 43.0& 54.6& 67.1& 64.3& 63.4\\
SSP& 2& 74.7& 58.8& 45.7& 55.6& 66.3& 63.3& 61.4\\
\hline\end{tabular}\\

\end{document}